\documentclass[aps,prd,
groupedaddress,amssymb,
showpacs,nofootinbib]{revtex4}

\usepackage{amsmath,amssymb}

\usepackage[usenames]{color}

\usepackage{graphicx}
\usepackage{mathptmx}
\usepackage{bm}
\usepackage{times}

\newcommand{\beq}{\begin{eqnarray}}
\newcommand{\eeq}{\end{eqnarray}}

\begin{document}

\onecolumngrid
\begin{flushright}
BCCUNY-HEP/09-02
\end{flushright}


\title{Longitudinal Rescaling and High-Energy Effective Actions}



\author{Peter \surname{Orland}}

\email{orland@nbi.dk}


\author{Jing \surname{Xiao}}

\email{xiao9304@hotmail.com}

\affiliation{1. Baruch College, The 
City University of New York, 17 Lexington Avenue, 
New 
York, NY 10010, U.S.A. }

\affiliation{2. The Graduate School and University Center, The City University of New York, 365 Fifth Avenue,
New York, NY 10016, U.S.A.}

\date{January 20, 2009}

\begin{abstract}
Under a longitudinal rescaling of coordinates 
$x^{0,3}\rightarrow \lambda x^{0,3}$, $\lambda\ll 1$, the classical QCD action simplifies 
dramatically. This is the high-energy limit, as
$\lambda \sim s^{-1/2}$, where
$s$ is the center-of-mass energy squared of a hadronic
collision. We find the quantum corrections to the rescaled action at one loop, in particular
finding the anomalous powers of $\lambda$ in this action, with $\lambda<1$. The
method is an integration over high-momentum components of the
gauge field. This is a Wilsonian renormalization procedure, and 
counterterms are needed to make the sharp-momentum cut-off
gauge invariant. Our
result for the quantum action is found, assuming 
$\vert \ln \lambda\vert \ll1$, which is essential for the validity
of perturbation theory. If $\lambda$ is sufficiently small (so that $\vert \ln \lambda\vert\gg1$), then the perturbative 
renormalization
group breaks down. This is due to uncontrollable fluctuations of
the longitudinal chromomagnetic field.

\end{abstract}


\maketitle

\section{Introduction}
\setcounter{equation}{0}
\renewcommand{\theequation}{1.\arabic{equation}}

Effective gauge-theory descriptions are a promising
approach to high-energy proton-proton collisions
\cite{Lipatov}, \cite{Kovchegov}, 
\cite{Verlinde-squared}, 
\cite{KirschLipSzyman}, \cite{mypaper},  and nuclear collisions
\cite{McLerranVenugopalan}, \cite{CGC}, \cite{Jalilian-Marian-et-al},
\cite{Jalilian-Marian-et-al-2},  \cite{Hatta-etal}.

The  
approximation of Verlinde and Verlinde \cite{Verlinde-squared} was to eliminate
some gauge-theory degrees of freedom through a longitudinal rescaling. These
authors argued that this rescaling yields the BFKL theory
\cite{BalitskiFadKurLip}. In particular, they
were able to re-derive the BFKL vertex and argued that
gluon Reggeization occurs. A similar
idea was incorporated by McLerran and Venugopalan 
\cite{McLerranVenugopalan} into a picture which came to be known as
the Color-Glass Condensate
\cite{CGC}. Longitudinal rescaling in Reference \cite{Verlinde-squared}
was done purely {\em classically}, by a simple change of variables in the action. After the
rescaling, quark and gluon matter travels primarily longitudinally. Most of the
energy is in the transverse color field strength, just as in a Weizsacker-Williams
shock wave. Effective actions incorporating such shock waves have been extensively
discussed by Lipatov \cite{Lip} and Balitsky \cite{Bal}

In Reference \cite{mypaper}, the cut-off rescaled theory was shown
to be completely integrable, massive and confining, in the high-energy limit. Our interest
here is to see whether this limit is justified.

In this paper, we
determine how the quantum 
action changes under longitudinal rescaling. We will
only consider the gluon field in our calculation. Quarks will be included in a later
publication.

The explicit rescaling of coordinates and gauge fields is
$x^{0,3}\rightarrow \lambda x^{0,3}$,
$x^{1,2}\rightarrow x^{1,2}$,
$A_{0,3}\rightarrow \lambda^{-1} A_{0,3}$, $A_{1,2}\rightarrow A_{1,2}$, where
$A_{\mu}=A_{\mu}^{a}t_{a}$, $a=1,\dots, N^{2}-1$
are SU($N$) Yang-Mills field. Sometimes we shall use $L$
as an abbreviation for the longitudinal Lorentz indices $0,3$ and $\perp$
as an abbreviation for the transverse Lorentz indices
$1,2$. We 
normalize ${\rm Tr}\,t_{a}t_{b}=\delta_{ab}$
and define ${\rm i}f_{ab}^{c}t_{c}=[t_{a},t_{b}]$. Since momentum components
transform as
$p_{L}\rightarrow \lambda^{-1}p_{L}$, $p_{\perp}\rightarrow p_{\perp}$, we can think of
the rescaling factor as $\lambda=\sqrt{s^{\prime}/s}$, where $s^{\prime}$ and
$s$ are the center-of-mass
energies squared, before and after the rescaling, respectively. To
describe extremely high energies, we would, in principle, take
$\lambda \ll 1$ \cite{Verlinde-squared}.

Perhaps a better motivation for this rescaling
is that transverse transport of glue is suppressed and
longitudinal transport is enhanced. This can be seen by perusing the
Hamiltonian. If the scale factor $\lambda$ is small, but
not zero, the resulting Hamiltonian has one extremely small coupling and
one extremely large coupling. The classically-rescaled action is 
\beq
S\!=\!\frac{1}{2g_{0}^{2}}\!\int d^{4}x 
{\rm Tr}\!\left(\! \lambda^{-2}F_{03}^{2}\!+\!\sum_{j=1}^{2}F_{0j}^{2}
\!-\!\sum_{j=1}^{2}F_{j3}^{2} -\lambda^{2} F_{12}^{2}\right)\!,
\label{action}
\eeq
where 
$F_{\mu \nu}=\partial_{\mu}A_{\nu}-\partial_{\nu}A_{\mu}-{\rm i}[A_{\mu},A_{\nu}]$.
The Hamiltonian in $A_{0}=0$ gauge is therefore
\beq
H= \int d^{3} x \left[\frac{g_{0}^{2}}{2}{\mathcal E}_{\perp}^{2}+
\frac{1}{2g_{0}^{2}}{\mathcal B}_{\perp}^{2}+
\lambda^{2}\left(\frac{g_{0}^{2}}{2}{\mathcal E}_{3}^{2} +
\frac{1}{2g_{0}^{2}}{\mathcal B}_{3}^{2}\right) \right], \label{ContHamiltonian}
\eeq
where 
the electric and magnetic fields are ${\mathcal E}_{i}=-{\rm i}\delta/\delta A_{i}$
and ${\mathcal B}_{i}=\epsilon^{ijk}(\partial_{j}A_{k}+A_{j}\times A_{k})$, respectively
and $(A_{j}\times A_{k})^{a}=f_{bc}^{a}A_{j}^{b}A_{k}^{c}$. Physical states $\Psi$ must 
satisfy Gauss's law
\beq
\left(\partial_{\perp}\cdot {\mathcal E}_{\perp}+\partial_{3}{\mathcal E}_{3}-\rho\right)\Psi=0\;,
\label{Gauss}
\eeq
where $\rho$ is the quark color-charge density. If the term of order $\lambda^{2}$ is 
neglected, all the energy is contained in the 
transverse electric and magnetic fields. Chromo-electromagnetic 
waves can only move longitudinally. This is most easily seen in an axial gauge $A_{3}=0$,
in which case the $\lambda=0$ Hamiltonian contains no transverse derivatives 
\cite{mypaper}.

As we mentioned above, the longitudinal rescaling considered above is
classical. In a fully-quantized Yang-Mills theory, the rescaled action is not as simple as
(\ref{action}). In the quantum case, all the coefficients of
the field strength-squared terms must be rescaled. Furthermore, these coefficients 
are not simply multiplied by integer powers of $\lambda$; anomalous dimensions
are present.

The rescaling is done for the quantized Yang-Mills theory in two steps. First a 
Wilson-style renormalization \cite{WK}
from an isotropic to an anisotropic cut-off is performed. Second, the
longitudinal rescaling discussed above is done to restore the isotropy of the cut-off. One
way to visualize this procedure is to imagine a lattice ultraviolet cut-off, with
lattice spacing $a$. The lattice rescaling procedure is illustrated
in Fig. 1. Degrees of freedom are thinned out
by a Kadanoff or ``block-spin" transformation, which changes the lattice spacing in the longitudinal
directions to $a/\lambda$, while leaving the lattice spacing in the transverse directions
unchanged. After this reduction of degrees of freedom, the 
entire lattice is rescaled longitudinally, so that
the lattice spacing in the direction of any coordinate axis has the original value $a$. 

Some papers on anisotropic renormalization were written 
\cite{Aref'evaVolovich}, not long after 
References \cite{Verlinde-squared}, \cite{McLerranVenugopalan} appeared. Perturbative renormalization of the Yang-Mills field is not performed in
these papers.

The Wick-rotated Yang-Mills
theory is defined by introducing the functional integral $\int \exp-S$, where $S$ is the action, with
an ultraviolet cut-off $\Lambda$ on the variables of 
integration, namely the gauge field $A_{\mu}(x)$ (we do not include 
quark fields in this paper). The cut-off is introduced by requiring that the Fourier
components of these fields, which are functions of Euclidean four-momentum $p$, vanish
for $p^{2}>\Lambda^{2}$. This sharp momentum cut-off breaks gauge invariance, meaning that
counterterms restoring this invariance are necessary. We denote the two components of
longitudinal momenta by $p_{L}=(p_{0},p_{3})$
and the two components of transverse momenta by $p_{\perp}=(p_{1},p_{2})$. 

We first isolate the degrees of freedom depending on momenta satisfying 
${\tilde b}p_{L}^{2}+p_{\perp}^{2}>{\tilde \Lambda^{2}}$, where $\Lambda\ge {\tilde \Lambda}$,
${\tilde b}\ge 1$, and integrate these 
out of the functional integral. This yields a new functional integral whose action has
new couplings, but with an ellipsoidal cut-off, with the remaining degrees of freedom
vanishing unless ${\tilde b}p_{L}^{2}+p_{\perp}^{2}< {\tilde \Lambda^{2}}$. The different
coefficients
of the field-strength-squared component are rescaled differently. Finally, we rescale 
$p_{L}\rightarrow \lambda^{-1}p_{L}$ and $p_{\perp}\rightarrow p_{\perp}$. We identify
$\lambda^{-2}={\tilde b}$. The ultraviolet regularization is once again isotropic, with
components vanishing unless $p^{2}=p_{L}^{2}+p_{\perp}^{2}<{\tilde \Lambda}^{2}$. As a
result, the 
different coefficients
of the field-strength-squared components are rescaled again, yielding the final form
of the action. 

It is possible to assume that ${\tilde \Lambda}=\Lambda$. In that case, we integrate out
all the degrees of freedom in the original momentum sphere, except for those
in an ellipsoid, whose
two major axes are equal to the diameter of the sphere. It is illustrative, however, to 
consider the more general case of ${\tilde \Lambda}\le 
\Lambda$.

The plan of the paper is as follows. In the next section, we discuss generally how the
Wilson renormalization for an SU($N$) Yang-Mills theory is carried out. The isotropic
case is briefly reviewed in Section 3. The integration from a spherical cut-off to
an ellipsoidal cut-off is explained in Section 4. This result is then used to
find the effect of a longitudinal rescaling on the Yang-Mills action in Section 5. We 
touch upon the utility of effective actions for
high-energy collisions, in the light of our results, in Section 
6. In the last section, we mention some 
calculations which should be done, in the near future.

\section{Renormalization of QCD with a momentum cut-off: General Considerations}
\setcounter{equation}{0}
\renewcommand{\theequation}{2.\arabic{equation}}

In this section, we review how the QCD action changes if we integrate, to one loop, 
from one sharp momentum cut-off 
to a smaller sharp momentum
cut-off. For readers not already familiar with this method, a discussion can be found in
Reference \cite{PolyakovBook}. The techniques do not differ appreciably from
those for the background-field calculation of the effective action.

First we Wick rotate the Yang-Mills theory to obtain the standard Euclidean
metric. We choose $\Lambda$ and $\tilde \Lambda$ to be real positive numbers with units of
$cm^{-1}$ and $b$ and $\tilde b$ to be two dimensionless real numbers, such that
$b\ge 1$ and ${\tilde b}\ge 1$. We require furthermore that $\Lambda>{\tilde \Lambda}$
and that $\Lambda^{2}/b \ge {\tilde \Lambda}^{2}/{\tilde b}$.  We  
define the region of momentum space $\mathbb P$ to be the set of points $p$, such that
$bp_{L}^{2}+p_{\perp}^{2}<\Lambda^{2}$. We define the region ${\tilde {\mathbb P}}$ to
be the
set of points $p$, such that ${\tilde b}p_{L}^{2}+p_{\perp}^{2}<{\tilde \Lambda}^{2}$. Finally,
we define $\mathbb S$ to be the Wilsonian 
``onion skin" ${\mathbb S}={\mathbb P}-{\tilde{\mathbb P}}$.

The functional integral we consider is
\beq
Z_{\Lambda}=\int \left[ \prod_{p\in {\mathbb P}} dA(p)\right] \exp -S ,\;\;\;
S=\int d^{4}x \frac{1}{4g_{0}^{2}}{\rm Tr}\; F_{\mu \nu}F^{\mu \nu}+S_{c.t., \Lambda,b}\, ,
\label{cut-off-FI-1}
\eeq
where 
$S_{c.t., \Lambda,b}$ contains counterterms, needed to maintain gauge invariance with the 
sharp-momentum cut-off $\Lambda$, and anisotropy parameter $b$.

The restriction
on the measure of integration in (\ref{cut-off-FI-1}) means that the gauge
field has the Fourier
transform
\beq
A_{\mu}(x)=\int_{\mathbb P} \frac{d^{4}p}{(2\pi)^{4}} \;A_{\mu}(p) \,\,e^{-{\rm i} p\cdot x}\;. \nonumber
\eeq

We split the field $A_{\mu}$ into slow parts ${\tilde A}_{\mu}$, and fast parts $a_{\mu}$, defined
by
\beq
{\tilde A}_{\mu}(x)=
\int_{\tilde {\mathbb P}} \frac{d^{4}p}{(2\pi)^{4}}\; A_{\mu}(p) \,\,e^{-{\rm i} p\cdot x} \;,
\;\;
a_{\mu}(x)=
\int_{{\mathbb S}} 
\frac{d^{4}p}{(2\pi)^{4}} \;A_{\mu}(p)\,\, e^{-{\rm i} p\cdot x} \;, \nonumber
\eeq
so that $A_{\mu}(x)={\tilde A}_{\mu}(x)+a_{\mu}(x)$. This can also be written in momentum
space: $A_{\mu}(p)={\tilde A}_{\mu}(p)+a_{\mu}(p)$, by defining
\beq
{\tilde A}_{\mu}(p)=\left\{  \begin{array}{cc} A_{\mu}(p), & p\in {\tilde {\mathbb P}},\\
0,& p\in {\mathbb S} \end{array} \right.,\;\;\;a_{\mu}(p)=\left\{  \begin{array}{cc} 0, & p\in {\tilde {\mathbb P}},\\
A_{\mu}(p),& p\in {\mathbb S} \end{array} \right. \;. 
\eeq

We shall integrate out the fast components
$a_{\mu}$,
of the field to obtain
\beq
Z_{\Lambda}=e^{-f}Z_{{\tilde \Lambda}}\;,\;\;
Z_{\tilde \Lambda}=\int \left[ \prod_{p\in {\tilde {\mathbb P}}} dA(p)\right] 
\exp -{\tilde S},
\;\;
{\tilde S}=\int d^{4}x\, \frac{1}{4{\tilde g}_{0}^{2}}\,\,{\rm Tr} \;{\tilde F}_{\mu \nu}{\tilde F}^{\mu \nu}
+S_{c.t., {\tilde \Lambda}, {\tilde b}}\;,
\label{cut-off-FI-2}
\eeq
where $f$ is an unimportant ground-state-energy renormalization, ${\tilde g}_{0}$ is the coupling
at the new cut-off ${\tilde \Lambda}$, ${\tilde b}$,
${\tilde F}_{\mu \nu}=\partial_{\mu}{\tilde A}_{\nu}-\partial_{\nu}{\tilde A}_{\mu}-{\rm i}[{\tilde A}_{\mu},
{\tilde A}_{\nu}]$, and $S_{c.t., {\tilde \Lambda},{\tilde b}}$ contains the counterterms needed to restore gauge
invariance with the new cut-off.

To integrate out the fast gauge field, yielding the new action in (\ref{cut-off-FI-2}), we expand the
original action in terms of this field to quadratic order:
\beq
S=\frac{1}{4g_{0}^{2}}\int d^{4}x\; {\rm Tr}\left\{
{\tilde F}_{\mu \nu}{\tilde F}^{\mu \nu}
-4[{\tilde D}_{\mu}, {\tilde F}^{\mu \nu}  ]a_{\nu}
+([{\tilde D}_{\mu},a_{\nu}]-[{\tilde D}_{\nu},a_{\mu}])
([{\tilde D}^{\mu},a^{\nu}]-[{\tilde D}^{\nu},a^{\mu}])
-2{\rm i}  {\tilde F}^{\mu \nu}[a_{\mu},a_{\nu}]
\right\}\;, \label{quadratic-expansion} 
\eeq
where ${\tilde D}_{\mu}=\partial_{\mu}-{\rm i}{\tilde A}_{\mu}$ is the covariant derivative determined by
the slow gauge field. 

The action is invariant under the gauge transformation of the
fast field:
\beq
{\tilde A}_{\mu}\rightarrow {\tilde A}_{\mu}\;,\;\;
a_{\mu}\rightarrow a_{\mu}+[{\tilde D}_{\mu}-{\rm i}a_{\mu},\omega]\;. 
\nonumber
\eeq
Variations $\delta a_{\mu}$ orthogonal to these gauge transformation satisfy 
$[{\tilde D}_{\mu},\delta a_{\mu}]=0$. We can add with impunity the term
$\frac{1}{2g_{0}^{2}}\int d^{4}x{\rm Tr}[{\tilde D}_{\mu},a_{\mu}]^{2}$ to the action.

Notice that there is a linear term in $a_{\mu}$ in the action (\ref{quadratic-expansion}). Once we integrate out the fast field, the only result of this term will be to induce terms of order
$[{\tilde D}_{\mu}, {\tilde F}^{\mu \nu}  ]^{2}$ in ${\tilde S}$. These terms will be of dimension
greater than four or nonlocal, so we ignore them, as they will be irrelevant.  We can
thereby replace (\ref{quadratic-expansion}) with
\beq
S=\frac{1}{4g_{0}^{2}}\int d^{4}x\; {\rm Tr}
{\tilde F}_{\mu \nu}{\tilde F}^{\mu \nu}
+\frac{1}{2g_{0}^{2}}\int d^{4} x \left([{\tilde D}_{\mu},a_{\nu}]
[{\tilde D}^{\mu},a^{\nu}]
-2{\rm i}  {\tilde F}^{\mu \nu}[a_{\mu},a_{\nu}]
\right) \;, 
\nonumber
\eeq
In terms of coefficients of the generators $t_{b}$, $b=1,\dots,N^{2}-1$, this expression may be
written as
\beq
S=\frac{1}{4g_{0}^{2}}\int d^{4}x\; 
{\tilde F}^{b}_{\mu \nu}{\tilde F}_{b}^{\mu \nu}+
S_{\rm O}+S_{\rm I}+S_{\rm II}\;, 
\nonumber
\eeq
where
\beq
S_{\rm O}=\frac{1}{2g_{0}^{2}}\int_{\mathbb S} \frac{d^{4}q}{(2\pi)^{4}} \;q^{2}\;a^{b}_{\mu}(-q)
a_{b}^{\mu}(q)\;,\label{SO}
\eeq
\beq
S_{\rm I}&=&\frac{{\rm i}}{g_{0}^{2}}\int_{\mathbb S} \frac{d^{4}q}{(2\pi)^{4}} 
\int_{\tilde {\mathbb P}}\frac{d^{4}p}{(2\pi)^{4}} 
q^{\mu} f_{bcd} a^{b}_{\nu}(q)\,{\tilde A}^{c}_{\mu}(p)\,a^{d}_{\nu}(-q-p) \nonumber \\
&+&\frac{1}{2g_{0}^{2}}\int_{\mathbb S} \frac{d^{4}q}{(2\pi)^{4}}\int_{\tilde {\mathbb P}}\frac{d^{4}p}{(2\pi)^{4}}
\int_{\tilde {\mathbb P}}\frac{d^{4}l}{(2\pi)^{4}} f_{bcd} f_{bfg}\, a^{d}_{\nu}(q)\,{\tilde A}^{c}_{\mu}(p)
{\tilde A}^{f}_{\mu}(l)\,a^{g}_{\nu}(-q-p)
\;,
\label{SI}
\eeq
and
\beq
S_{\rm II}=\frac{1}{2g_{0}^{2}}\int_{\mathbb S} \frac{d^{4}q}{(2\pi)^{4}}\int_{\tilde {\mathbb P}}\frac{d^{4}p}{(2\pi)^{4}}
f_{bcd} \,a^{b}_{\mu}(q) {\tilde F}_{\mu\nu}^{c}(p)a^{d}_{\nu}(-p-q)
\;.
\label{SII}
\eeq
The gluon propagator is given by the expression for $S_{\rm O}$ in (\ref{SO}) as
\beq
\langle a^{b}_{\mu}(q) a^{c}_{\nu}(p) \rangle
=g_{0}^{2} \delta^{bc}\delta_{\mu \nu} \delta^{4}(q+p)q^{-2}\;.
\label{propagator}
\eeq
We define the meaning of brackets $\langle W\rangle$, around any quantity $W$ to be
the expectation value of $W$ with respect to the measure ${\mathcal N}\exp-S_{\rm O}$, 
where $\mathcal N$ is chosen so that $\langle 1 \rangle=1$. 

One more term must be included in the action, which depends on the anticommuting ghost
fields $G^{b}_{\mu}(x)$, $H^{b}_{\mu}(x)$, associated 
with the gauge fixing of $a^{b}_{\mu}(x)$. The ghost action is
\beq
S_{\rm ghost}&=&\frac{{\rm i}}{g_{0}^{2}}\int_{\mathbb S} \frac{d^{4}q}{(2\pi)^{4}} 
\int_{\tilde {\mathbb P}}\frac{d^{4}p}{(2\pi)^{4}} 
q^{\mu} f_{bcd} G^{b}(q)\,{\tilde A}^{c}_{\mu}(p)\,H^{d}(-q-p) \nonumber \\
&+&\frac{1}{2g_{0}^{2}}\int_{\mathbb S}
 \frac{d^{4}q}{(2\pi)^{4}}\int_{\tilde {\mathbb P}}\frac{d^{4}p}{(2\pi)^{4}}
\int_{\tilde {\mathbb P}}\frac{d^{4}l}{(2\pi)^{4}} 
f_{bcd} f_{bfg}\, G^{d}(q)\,{\tilde A}^{c}_{\mu}(p){\tilde A}^{f}_{\mu}(l)\,
H^{g}(-q-p)
\;,
\nonumber
\eeq
which is similar to $S_{\rm I}$, except that
the fast vector gauge field has been replaced by the scalar ghost fields. Integration over the
ghost fields eliminates two of the four spin degrees of freedom of the fast gauge field.

To 
integrate out the fast gauge field and its associated ghost fields, we use the connected-graph
expansion for the expectation value of the exponential of minus a quantity $R$:
\beq
\langle e^{-R}\rangle=\exp\left[-\langle R\rangle+\frac{1}{2!}(\langle R^{2}\rangle-\langle R\rangle^{2})
-\frac{1}{3!}\left( \langle R^{3}\rangle-3\langle R^{3}\rangle\langle R\rangle
+2\langle R\rangle^{3}\right) +\cdots
\right]\;.
\nonumber
\eeq
Applying this expansion to second order, we find
\beq
\exp-{\tilde S}&=&\exp\left(-\frac{1}{4g_{0}^{2}}\int d^{4}x\; 
{\tilde F}^{b}_{\mu \nu}{\tilde F}_{b}^{\mu \nu}\right)
\left\langle \exp\left(-\frac{1}{2}S_{\rm I}-S_{\rm II}\right) \right\rangle
\label{int-out} \nonumber \\
&\approx&\exp\left[-\frac{1}{4g_{0}^{2}}\int d^{4}x\; 
{\tilde F}^{b}_{\mu \nu}{\tilde F}_{b}^{\mu \nu} \right]
\exp \left[
-\frac{1}{2}\langle S_{\rm I}\rangle
+\frac{1}{4}(\langle S_{\rm I}^{2}\rangle-\langle S_{\rm I}\rangle^{2})
+\frac{1}{2}(\langle S_{\rm II}^{2}\rangle-\langle S_{\rm II}\rangle^{2}) 
\right] \;.\label{one-loop}
\eeq
We remark briefly on the coefficients in the last 
exponential in (\ref{one-loop}). The coefficient of $\langle S_{\rm I}\rangle$ has a contribution $-1$ from
a fast gluon loop and $1/2$ from a fast ghost loop. The coefficient of 
$\langle S_{\rm I}^{2}\rangle-\langle S_{\rm I}\rangle^{2}$ has
a contribution $1/2$ from a fast gluon loop and $-1/4$ from a fast 
ghost loop. The 
coefficient of $\langle S_{\rm II}^{2}\rangle-\langle S_{\rm II}\rangle^{2}$ has
no ghost contribution. Other 
terms in the exponential of the same order vanish 
upon contraction of group indices.

The terms in the new action (\ref{one-loop}) are given by
\beq
\frac{1}{2}\langle S_{\rm I}\rangle\!\!&\!\!-\!\!&\!\!\frac{1}{4}(\langle S_{\rm I}^{2}\rangle-\langle S_{\rm I}\rangle^{2}) =\frac{C_{N}}{4}\int_{\tilde {\mathbb P}} \frac{d^{4}p}{(2\pi)^{4}}
{\tilde A}^{b}_{\mu}(-p){\tilde A}^{b}_{\nu}(p) P_{\mu \nu}(p)\;, \nonumber \\
P_{\mu \nu}(p)\!\!&\!\!=\!\!&\!\! \int_{{\mathbb S}}\frac{d^{4}q}{(2\pi)^{4}}\left[-\frac{q_{\mu}(p_{\nu}+2q_{\nu})}{4q^{2}(q+p)^{2}}+\frac{\delta_{\mu \nu}}{4q^{2}}\right]  \label{pol-tensor}
\;,
\eeq
where $C_{N}$ is the Casimir of SU($N$), defined by $f^{bcd}f^{hcd}=C_{N}\delta^{bh}$,
and
\beq
-\frac{1}{2}(\langle S_{\rm II}^{2}\rangle-\langle S_{\rm II}\rangle^{2}) 
=-\frac{C_{N}}{2}\int_{\tilde {\mathbb P}} \frac{d^{4}p}{(2\pi)^{4}} 
{\tilde F}^{b}_{\mu \nu}(-p)
{\tilde F}^{b}_{\mu \nu}(p)
\int_{{\mathbb S}} \frac{d^{4}q}{(2\pi)^{4}} \frac{1}{q^{2}(p+q)^{2}}  \;. \label{F-F}
\eeq

The remaining work to be done is
to evaluate integrals in (\ref{pol-tensor}) and (\ref{F-F}). 

Notice that if the integral $I(p)$ is defined by 
\beq
I(p)=\int_{ {\mathbb S}} \frac{d^{4}q}{(2\pi)^{4}}\frac{p_{\alpha}+2q_{\alpha}}{q^{2}(q+p)^{2}} \;,
\nonumber
\eeq
then $I(p)+I(-p)=0$. We can see this by changing the sign of $q$ in the integration. Hence we can
replace the polarization tensor $P_{\mu\nu}(p)$ in (\ref{pol-tensor}) by the manifestly symmetric 
form $\Pi_{\mu\nu}(p)$:
\beq
\frac{1}{2}\langle S_{\rm I}\rangle\!\!&\!\!-\!\!&\!\!\frac{1}{4}(\langle S_{\rm I}^{2}\rangle-\langle S_{\rm I}\rangle^{2}) =C_{N}\int_{\tilde {\mathbb P}} \frac{d^{4}p}{(2\pi)^{4}}\;
{\tilde A}^{b}_{\mu}(-p){\tilde A}^{b}_{\nu}(p)\; \Pi_{\mu \nu}(p)\;, \nonumber \\
\Pi_{\mu \nu}(p)\!\!&\!\!=\!\!&\!\! \int_{ {\mathbb S}}\frac{d^{4}q}{(2\pi)^{4}}\left[-\frac{
(p_{\mu}+2q_{\mu})(p_{\nu}+2q_{\nu})}{8q^{2}(q+p)^{2}}+\frac{\delta_{\mu \nu}}{4q^{2}}\right]  
\label{sym-pol-tensor}
\;.
\eeq
As it is now defined, the polarization tensor is symmetric, but breaks gauge invariance. This
is because at this order in the loop expansion, $p_{\mu}\Pi_{\mu\nu}(p)\neq 0$. The reason for this
is clear; gauge symmetry is explicitly broken by sharp-momentum cut-offs. The purpose of the 
counterterms $S_{c.t., \Lambda,b}$ and $S_{c.t., {\tilde \Lambda},{\tilde b}}$
in (\ref{cut-off-FI-1}) and (\ref{cut-off-FI-2}), respectively, is to restore this symmetry.

\section{Renormalization of QCD with a momentum cut-off: the spherical case}
\setcounter{equation}{0}
\renewcommand{\theequation}{3.\arabic{equation}}

Next we present the results of the one-loop calculation presented in the
last section for spherical cut-offs, {\em i.e.} $b={\tilde b}=1$. Absolutely nothing new will be found in this
section. Our only reason for
discussing the spherical case is that it is a
serviceable template for the more complicated ellipsoidal case.

Let us first evaluate $\Pi_{\mu \nu}(p)$ in (\ref{sym-pol-tensor}), segregating it into a gauge-invariant
part and a non-gauge-invariant part. At $p=0$,
\beq
\Pi_{\mu\nu}(0)=\int_{\mathbb S}\frac{d^{4}q}{(2\pi)^{4}}\left[-\frac{q_{\mu}q_{\nu}}{2(q^{2})^{2}} 
+\frac{\delta_{\mu\nu}}{4q^{2}}\right]\;. \nonumber
\eeq 
If we change the sign of one component only of $q$, {\em e.g.} $q_{0}\rightarrow -q_{0}$, 
$q_{\rm i}\rightarrow q_{\rm i}$, i$=1,2,3$, the first term of the integrand changes sign for $\mu=0$ and $\nu=$i. Hence $\Pi_{\mu\nu}(0)$ vanishes unless $\mu=\nu$. Thus
\beq
\Pi_{\mu\nu}(0)=\frac{1}{8}\int_{\mathbb S}\frac{d^{4}q}{(2\pi)^{4}}
\frac{\delta_{\mu\nu}}{q^{2}} =\frac{1}{128\pi^{2}}(\Lambda^{2}-{\tilde \Lambda}^{2})\delta_{\mu\nu}
\;. 
\nonumber
\eeq 
Writing $\Pi_{\mu\nu}(p)={\hat \Pi}_{\mu \nu}(p)+\Pi_{\mu\nu}(0)$, we find
\beq
{\hat \Pi}_{\mu \nu}(p)\!\!&\!\!=\!\!&\!\! \int_{ {\mathbb S}}\frac{d^{4}q}{(2\pi)^{4}}\left[-\frac{
(p_{\mu}+2q_{\mu})(p_{\nu}+2q_{\nu})}{8q^{2}(q+p)^{2}}+\frac{\delta_{\mu \nu}}{8q^{2}}\right]  
\nonumber
\;.
\eeq
If we subtract the polarization tensor
at zero momentum by a counterterms of identical form at each scale, or in other words
\beq
S_{c.t., \Lambda}=-\frac{\Lambda^{2}}{128\pi^{2}}\int d^{4}x \; \;A^{2}\;,\;\;
S_{c.t., {\tilde \Lambda}}=-\frac{{\tilde \Lambda}^{2}}{128\pi^{2}}\int d^{4}x\; \;{\tilde A}^{2}\;,
\label{spherical-counterterms}
\eeq
the result is gauge invariant, as we show below. 

Next we expand the polarization tensor ${\hat \Pi}_{\mu \nu}(p)$ in powers of $p$. The terms which
are more than quadratic order in $p$  have canonical dimension greater than four, so 
can be ignored in the new action. To this order,
\beq
{\hat \Pi}_{\mu \nu}(p)=
\int_{\mathbb S}\frac{d^{4}q}{(2\pi)^{4}} \left[ \frac{p_{\mu}p_{\nu}+\delta_{\mu\nu}p^{2}}{8(q^{2})^{2}}
-\frac{2p_{\alpha}p_{\beta}q_{\alpha}q_{\beta}q_{\mu}q_{\nu}}{(q^{2})^{4}}\right]+\cdots
\label{expanded-pol-tensor}
\eeq

The right-hand side of  (\ref{expanded-pol-tensor}) is readily 
evaluated using Euclidean O($4$) symmetry: we
emphasize this point, because in the aspherical case, we will not have invariance under O($4$), but
under its subgroup ${\rm O}(2)\times {\rm O}(2)$.  Exploiting this symmetry, we 
write the  nontrivial 
tensor integral in (\ref{expanded-pol-tensor}) in terms of a scalar integral:
\beq
\int_{\mathbb S}\frac{d^{4}q}{(2\pi)^{4}} \frac{q_{\alpha}q_{\beta}q_{\mu}q_{\nu}}{(q^{2})^{4}}
=\frac{1}{24}\int_{\mathbb S}\frac{d^{4}q}{(2\pi)^{4}} \frac{1}{q^{2}}
\left( \delta_{\alpha \beta}\delta_{\mu \nu}+
\delta_{\alpha \nu}\delta_{\mu \beta}+\delta_{\alpha \mu}\delta_{\beta \nu}
\right)\;.   
\nonumber
\eeq
The polarization tensor is therefore
\beq
{\hat {\Pi}}_{\mu\nu}(p)=\frac{1}{192\pi^{2}}\ln \frac{\Lambda}{\tilde \Lambda}\; (\delta_{\mu\nu}
-p_{\mu}p_{\nu})+\cdots\;. 
\label{polarization-part}
\eeq
Gauge invariance is satisfied to this order of $p$, {\em i.e.}
$p^{\mu}{\hat \Pi}_{\mu \nu}(p)=0$. 

Next we turn to (\ref{F-F}). As before, the terms of dimension higher than four can be
dropped, by expanding the integral over $\mathbb S$ in powers of $p$:
\beq
-\frac{1}{2}(\langle S_{\rm II}^{2}\rangle-\langle S_{\rm II}\rangle^{2}) 
&=&-\frac{C_{N}}{2}\int_{\tilde {\mathbb P}} \frac{d^{4}p}{(2\pi)^{4}} 
{\tilde F}^{b}_{\mu \nu}(-p)
{\tilde F}^{b}_{\mu \nu}(p)
\int_{{\mathbb S}} \frac{d^{4}q}{(2\pi)^{4}} \frac{1}{(q^{2})^{2}}
+\cdots  \nonumber \\
&=&-\frac{C_{N}}{16\pi^{2}}\;\ln \frac{\Lambda}{{\tilde \Lambda}}\;\int_{\tilde {\mathbb P}} 
\frac{d^{4}p}{(2\pi)^{4}} 
{\tilde F}^{b}_{\mu \nu}(-p)
{\tilde F}^{b}_{\mu \nu}(p)
+\cdots 
\;. \label{expanded-F-F}
\eeq

Putting together (\ref{sym-pol-tensor}), (\ref{spherical-counterterms}), (\ref{polarization-part}) and 
(\ref{expanded-F-F}) gives the standard result for the new coupling ${\tilde g}_{0}$ in (\ref{cut-off-FI-2}):
\beq
\frac{1}{{\tilde g}_{0}^{2}}=
\frac{1}{g_{0}^{2}}
-\frac{C_{N}}{8\pi^{2}}\; \ln\frac{\Lambda}{{\tilde \Lambda}}
+\frac{1}{12}\frac{C_{N}}{8\pi^{2}}\; \ln\frac{\Lambda}{{\tilde \Lambda}}
=\frac{1}{g_{0}^{2}}-
\frac{11 \,C_{N}}{96\pi^{2}} \ln \frac{\Lambda}{{\tilde \Lambda}}\;.
\label{asymptotic-freedom}
\eeq

\section{Renormalization of QCD with a momentum cut-off: the ellipsoidal case}
\setcounter{equation}{0}
\renewcommand{\theequation}{4.\arabic{equation}}

In the general case of ellipsoidal cut-offs, integration over the region $\mathbb S$ is
done by the change of variables, from $q_{\mu}$ to two angles $\theta$ and $\phi$, 
and two variables with dimensions of momentum squared, $u$ and $w$. The relation
between the old and new variables is
\beq
q_{1}=\sqrt{u}\, \cos\theta,\; q_{2}=\sqrt{u}\, \sin\theta,\;
q_{3}={\sqrt{w-u}}\, \cos\phi,\; q_{0}={\sqrt{w-u}}\, \sin\phi
\label{change-of-var}
\eeq
(note that $u=q_{\perp}^{2}$ and $w-u=q_{L}^{2}$), which gives {\large
\beq
\int_{\mathbb S}d^{4}q=\frac{1}{4}\int_{0}^{2\pi}d\theta\int_{0}^{2\pi}d\phi
\left[ 
\int_{0}^{{\tilde \Lambda}^{2}} du
\int_{{\tilde b}^{-1}{\tilde \Lambda}^{2}+(1-{\tilde b}^{-1})u}^{b^{-1}\Lambda^{2}+(1-b^{-1})u}
dw
+\int_{{\tilde \Lambda}^{2}}^{\Lambda^{2}} du
\int_{u}^{b^{-1}\Lambda^{2}+(1-b^{-1})u}dw          
\right] \;. \label{integration-measure}
\eeq}
The ${\rm O}(2)\times{\rm O}(2)$ symmetry group is generated by translations of the angles
$\theta\rightarrow \theta+d\theta$ and $\phi\rightarrow \phi+d\phi$.

The
polarization tensor $\Pi_{\mu\nu}(p)$ in (\ref{sym-pol-tensor}), expanded to second
order in $p_{\alpha}$ may be written as the sum of six terms: 
\beq
\Pi_{\mu \nu}(p)=\Pi_{\mu \nu}^{1}(p)+\Pi_{\mu \nu}^{2}(p)+\Pi_{\mu \nu}^{3}(p)+
\Pi_{\mu \nu}^{4}(p)+\Pi_{\mu \nu}^{5}(p)+\Pi_{\mu \nu}^{6}(p)\;,
\nonumber
\eeq
where 
\beq
\Pi_{\mu \nu}^{1}(p)&=&\frac{\delta_{\mu\nu}}{4} 
\int_{ {\mathbb S}}\frac{d^{4}q}{(2\pi)^{4}} \frac{1}{q^{2}}\;,\;\;
\Pi_{\mu\nu}^{2}(p)\,=\,-\frac{1}{2} \int_{\mathbb S}\frac{d^{4}q}{(2\pi)^{4}} \frac{q_{\mu}q_{\nu}}{(q^{2})^{2}},
\nonumber \\
\Pi_{\mu\nu}^{3}(p)&=&\frac{p_{\mu}p_{\alpha}}{2}\int_{ {\mathbb S}}\frac{d^{4}q}{(2\pi)^{4}} 
\frac{q_{\nu}q_{\alpha}}{(q^{2})^{3}}
+\frac{p_{\nu}p_{\alpha}}{2}\int_{ {\mathbb S}}\frac{d^{4}q}{(2\pi)^{4}} 
\frac{q_{\mu}q_{\alpha}}{(q^{2})^{3}}\;,\nonumber \\
\Pi_{\mu\nu}^{4}(p)&=&-\frac{p_{\mu}p_{\nu}}{8}\int_{ {\mathbb S}}\frac{d^{4}q}{(2\pi)^{4}} 
\frac{1}{(q^{2})^{2}}\;,\;\;
\Pi_{\mu\nu}^{5}(p)\,=\,\frac{p^{2}}{2}\int_{ {\mathbb S}}\frac{d^{4}q}{(2\pi)^{4}} 
\frac{q_{\mu}q_{\nu}}{(q^{2})^{3}}\;, \nonumber \\
\Pi_{\mu\nu}^{6}(p)&=&-2p_{\alpha}p_{\beta}I_{\alpha \beta \mu \nu}^{6}(p)\;,\;{\rm where}\;\,
I_{\alpha \beta \mu \nu}^{6}(p)=
\int_{ {\mathbb S}}\frac{d^{4}q}{(2\pi)^{4}} 
\frac{q_{\alpha}q_{\beta}q_{\mu}q_{\nu}}{(q^{2})^{4}} \;. \label{six-pol-terms}
\eeq

We next evaluate each of the six terms of the polarization tensor 
(\ref{six-pol-terms}). This is done using the integration 
(\ref{integration-measure}) over the variables (\ref{change-of-var}) which is
tedious, though not difficult. Since the integrals are invariant under 
${\rm O}(2) \times {\rm O}(2)$, but 
not O($4$),
we introduce a bit of notation. We
assume the indices $C$ and $D$ take only the values $1$ and
$2$, and the indices $\Omega$ and $\Xi$ take only the values $3$ and $0$. As usual, the
indices $\mu$, $\nu$, etc., can take any of the four values $1$, $2$, $3$ and $0$. The 
results are
\beq
\Pi_{\mu \nu}^{1}(p)
=\frac{\delta_{\mu\nu}}{64\pi^{2}}\left(
\frac{ \Lambda^{2}\ln b}{b-1}
-\frac{ {\tilde \Lambda}^{2}\ln {\tilde b}}{{\tilde b}-1} \right)\;,
\label{pi-1}
\eeq
\beq
\Pi_{CD}^{2}(p)
\!\!&\!\!=\!\!&\!\!-\frac{\Lambda^{2} \delta_{CD}}{64\pi^{2}}
\left[ 1+\frac{b}{(b-1)^{2}} (1-b+\ln b)\right] 
+\frac{{\tilde \Lambda}^{2} \delta_{CD}}{64\pi^{2}}
\left[ 1+\frac{\tilde b}{({\tilde b}-1)^{2}} (1-{\tilde b}+\ln {\tilde b})\right] \;, \nonumber \\
\Pi_{{\Omega}{\Xi}}^{2}(p)
\!\!&\!\!=\!\!&\!\!-\frac{\Lambda^{2}\delta_{{\Omega}{\Xi}}}{64\pi^{2}} 
\left[ \frac{1}{b-1}-\frac{\ln b}{(b-1)^{2}} \right]
+\frac{{\tilde \Lambda}^{2}\delta_{{\Omega}{\Xi}}}{64\pi^{2}} \left[ \frac{1}{{\tilde b}-1}-
\frac{\ln {\tilde b}}{({\tilde b}-1)^{2}} \right] \;, \nonumber \\
\Pi_{C{\Omega}}^{2}(p)\!\!&\!\!=\!\!&\!\! \Pi_{{\Omega} C}^{2}(p)=0\;,
\label{pi-2}
\eeq
\beq
\Pi_{CD}^{3}(p)
\!\!&\!\!=\!\!&\!\! \frac{p_{C}p_{D}}{32\pi^{2}}\ln {\frac{\Lambda}{{\tilde \Lambda}}}
-\frac{p_{C}p_{D}}{64\pi^{2}}\left[
\frac{b\ln b}{(b-1)^{2}}-\frac{b}{b-1}\right]
+\frac{p_{C}p_{D}}{64\pi^{2}}\left[
\frac{{\tilde b}\ln {\tilde b}}{({\tilde b}-1)^{2}}-\frac{\tilde b}{{\tilde b}-1}\right]\;, \nonumber \\
\Pi_{\Omega \Xi}^{3}(p)
\!\!&\!\!=\!\!&\!\! \frac{p_{\Omega}p_{\Xi}}{32\pi^{2}}\ln \frac{\Lambda}{\tilde \Lambda}
-\frac{p_{\Omega}p_{\Xi}}{64\pi^{2}} \left[
\frac{2b\ln b}{b-1}-\frac{b\ln b}{(b-1)^{2}} +\frac{b}{b-1}
\right]
+\frac{p_{\Omega}p_{\Xi}}{64\pi^{2}} \left[
\frac{2{\tilde b}\ln {\tilde b}}{{\tilde b}-1}-
\frac{{\tilde b}\ln {\tilde b}}{({\tilde b}-1)^{2}} +\frac{\tilde b}{{\tilde b}-1}
\right]\;, \nonumber \\
\Pi_{C \Omega}^{3}(p)\!\!&\!\!=\!\!&\!\! \Pi_{\Omega C}^{3}(p)=
\frac{p_{C}p_{\Omega}}{32\pi^{2}}\ln \frac{\Lambda}{\tilde \Lambda}
-\frac{p_{C}p_{\Omega}}{64\pi^{2}} \frac{b \ln b}{b-1}
+\frac{p_{C}p_{\Omega}}{64\pi^{2}} \frac{{\tilde b} \ln {\tilde b}}{{\tilde b}-1}\;,
\label{pi-3}
\eeq
\beq
\Pi_{\mu \nu}^{4}(p)=-\frac{p_{\mu}p_{\nu}}{64\pi^{2}} \ln \frac{\Lambda}{\tilde \Lambda}
+\frac{p_{\mu}p_{\nu}}{128\pi^{2}}\left( \frac{b\ln b}{b-1} 
-\frac{{\tilde b}\ln {\tilde b}}{{\tilde b}-1} \right) \;, \label{pi-4}
\eeq
\beq
\Pi_{CD}^{5}(p)\!\!&\!\!=\!\!&\!\!\frac{p^{2}\delta_{CD}}{64\pi^{2}}
\ln \frac{\Lambda}{\tilde \Lambda}
-\frac{p^{2}\delta_{CD}}{128\pi^{2}}\left[ \frac{b \ln b}{(b-1)^{2}}-\frac{b}{b-1} \right]
+\frac{p^{2}\delta_{CD}}{128\pi^{2}}\left[ \frac{{\tilde b} \ln {\tilde b}}{({\tilde b}-1)^{2}}-
\frac{\tilde b}{{\tilde b}-1} \right]\;, \nonumber \\
\Pi_{\Omega \Xi}^{5}(p)\!\!&\!\!=\!\!&\!\!\frac{p^{2}\delta_{\Omega \Xi}}{64\pi^{2}}\ln 
\frac{\Lambda}{\tilde \Lambda}
-\frac{p^{2}\delta_{\Omega \Xi}}{128\pi^{2}}\left[ \frac{b (2b-3)\ln b}{(b-1)^{2}}+\frac{b}{b-1} \right]
+\frac{p^{2}\delta_{\Omega \Xi}}{128\pi^{2}}
\left[ \frac{{\tilde b} (2{\tilde b}-3)\ln {\tilde b}}{({\tilde b}-1)^{2}}+\frac{\tilde b}{{\tilde b}-1} \right]\;,
\nonumber \\
\Pi_{C\Omega}^{5}(p)\!\!&\!\!=\!\!&\!\! \Pi_{\Omega C}^{5}(p)=0\;,
\label{pi-5}
\eeq
and finally, we present the components of the tensor 
$I_{\alpha \beta \mu \nu}^{6}(p)$ (from 
which the components of $\Pi_{\mu \nu}^{6}(p)$
can be obtained)
\beq
I_{CCCC}^{6}(p)
\!\!&\!\!=\!\!&\!\!\frac{1}{64\pi^{2}}\ln \frac{\Lambda}{\tilde \Lambda}
-\frac{b^{3}}{128\pi^{2}(b-1)^{3}}\left[
\ln b -\frac{2(b-1)}{b}+\frac{b^{2}-1}{2b^{2}}
\right]    \nonumber \\
\!\!&\!\!+\!\!&\!\! \frac{{\tilde b}^{3}}{128\pi^{2}({\tilde b}-1)^{3}}\left[
\ln {\tilde b} -\frac{2({\tilde b}-1)}{\tilde b}+\frac{{\tilde b}^{2}-1}{2{\tilde b}^{2}}
\right]\;, \nonumber \\
I_{1122}^{6}(p)\!\!&\!\!=\!\!&\!\! \frac{1}{3}I_{CCCC}^{6}(p)\;, \nonumber \\
I_{\Omega \Omega \Omega \Omega}^{6}(p) \!\!&\!\!=\!\!&\!\! 
\frac{1}{64\pi^{2}}\ln \frac{\Lambda}{\tilde \Lambda}
-\frac{1}{64\pi^{2}(b-1)^{3}}\left[
\ln b -2(b-1)+\frac{b^{2}-1}{2} \right] \nonumber \\
\!\!&\!\!+\!\!&\!\! \frac{1}{64\pi^{2}({\tilde b}-1)^{3}}\left[
\ln {\tilde b} -2({\tilde b}-1)+\frac{{\tilde b}^{2}-1}{2} \right] \;, \nonumber \\
I_{0033}^{6}(p)\!\!&\!\!=\!\!&\!\! \frac{1}{3}I_{\Omega\Omega\Omega\Omega}^{6}(p)\;, \nonumber \\
I_{CC\Omega\Omega}^{6}\!\!&\!\!=\!\!&\!\! \frac{1}{192\pi^{2}}\ln \frac{\Lambda}{\tilde \Lambda}
-\frac{1}{384\pi^{2}} \left[
\frac{3b(2b-3)\ln b}{(b-1)^{2}}-\frac{2b^{3}\ln b}{(b-1)^{3}}+\frac{3b}{b-1}
+\frac{2b-1}{b}+\frac{b^{2}-1}{2b^{2}}
\right]  \nonumber \\
\!\!&\!\!+\!\!&\!\!\frac{1}{384\pi^{2}} \left[
\frac{3{\tilde b}(2{\tilde b}-3)\ln {\tilde b}}{({\tilde b}-1)^{2}}
-\frac{2{\tilde b}^{3}\ln {\tilde b}}{({\tilde b}-1)^{3}}+\frac{3{\tilde b}}{{\tilde b}-1}
+\frac{2{\tilde b}-1}{\tilde b}+\frac{{\tilde b}^{2}-1}{2{\tilde b}^{2}}
\right] \;.
\label{pi-6}
\eeq
All other nonvanishing components of $I_{\alpha \beta \mu \nu}^{6}(p)$ can be
obtained by permuting indices of those shown in (\ref{pi-6}).

Notice that $\Pi_{\mu \nu}^{j}(p)$, $j=1,\dots,6$ each change sign under the
interchange of $\Lambda$ and $b$ with $\tilde \Lambda$ and $\tilde b$, respectively. We can
eliminate $\Pi_{\mu \nu}^{1}(p)$ and $\Pi_{\mu\nu}^{2}(p)$ by a mass counterterm. The sum of
the other
pieces of the polarization tensor, $\sum_{j=3}^{6}\Pi_{\mu \nu}^{j}(p)$, reduces to the expression
in (\ref{polarization-part}) if $b={\tilde b}$; integrating degrees of freedom with momenta
between two similar ellipsoids yields the same result as integrating degrees of
freedom with momenta between two spheres.

Next we set $b=1$ and expand ${\tilde b}=1+\ln {\tilde b}+\cdots$. We drop the part of
the polarization tensor of order $(\ln {\tilde b})^{2}$. We write
the polarization tensor as matrix whose rows and columns are ordered by $1,2,3,0$.
After some work, we obtain 
\beq
\sum_{j=3}^{6}\Pi^{j}(p)\!\!&\!\!=\!\!&\!\! 
\frac{1}{192\pi^{2}}\ln \frac{\Lambda}{\tilde \Lambda}\; (1\!\!{\rm l}
-pp^{T})
\nonumber \\ 
&+&\frac{\ln {\tilde b}}{64 \pi^{2}}\left(   \begin{array}{cccc}
-\frac{3}{4}p_{1}^{2}-\frac{1}{6}p_{2}^{2}-\frac{13}{12}p_{L}^{2} &
-\frac{7}{12}p_{1}p_{2}& -\frac{7}{4}p_{1}p_{3}& -\frac{7}{4}p_{1}p_{0}\\ \\
-\frac{7}{12}p_{1}p_{2}& -\frac{3}{4}p_{2}^{2}-\frac{1}{6}p_{1}^{2}-\frac{13}{12}p_{L}^{2} &
-\frac{7}{4}p_{2}p_{3} & -\frac{7}{4} p_{2}p_{0} \\ \\
-\frac{7}{4}p_{1}p_{3} & -\frac{7}{4}p_{2}p_{3} &
\frac{7}{4}p_{3}^{2}+\frac{2}{3}p_{0}^{2}+\frac{1}{3}p_{\perp}^{2} &
\frac{13}{12}p_{3}p_{0} \\  \\
-\frac{7}{4}p_{1}p_{0} & -\frac{7}{4} p_{2}p_{0} & \frac{13}{12}p_{3}p_{0} &
\frac{2}{3}p_{3}^{2}+\frac{7}{4}p_{0}^{2}+\frac{1}{3}p_{\perp}^{2} 
\end{array}
\right)\;, \label{p-t}
\eeq
where $1\!\!{\rm l}$ is the four-by-four identity matrix and
the superscript $T$ denotes the transpose. The 
first term on the right-hand side of (\ref{p-t}) is the polarization tensor found in the
previous section (\ref{polarization-part}). The second term
does not depend on $\Lambda$ or $\tilde \Lambda$. Had we taken $b>1$, and
expanded $b=1+\ln b+\cdots$, the quantity
$\ln {\tilde b}$ in (\ref{p-t}) would have been $\ln ({\tilde b}/b)$. 

Notice that
the second term on the right-hand side of 
(\ref{p-t}) violates gauge invariance (multiplying the vector $p$
by the matrix in this term does not yield zero). Therefore, an additional counterterm is
necessary. The most general local action of dimension $4$, which is
quadratic in ${\tilde A}_{\mu}$ and which does not change under 
${\rm O}(2)\times{\rm O}(2)$ transformations and is gauge invariant to linear order
is
\beq
S_{\rm quad}=\int_{\tilde {\mathbb P}}\frac{d^{4}p}{(2\pi)^{4}}\;{\rm Tr}\;
{\tilde A}(-p)^{T}[a_{1}M_{1}(p)+a_{2}M_{2}(p)+a_{3}M_{3}(p)]
{\tilde A}(p)\;,
\nonumber
\eeq
where $a_{1}$, $a_{2}$ and $a_{3}$ are real coefficients and
\beq
M_{1}(p)&=&\left(   \begin{array}{cccc}
p_{2}^{2} & -p_{1}p_{2} &0 & 0 \\
-p_{1}p_{2}& p_{1}^{2} &0 & 0\\
0&0&0&0\\
0&0&0&0 \end{array} \right)\;,\;\;
M_{2}(p)=\left(   \begin{array}{cccc}
0&0&0&0\\
0&0&0&0 \\
0&0&p_{3}^{2} & -p_{3}p_{0}  \\
0& 0& -p_{3}p_{0}& p_{0}^{2}
\end{array} \right)\;,\nonumber \\
M_{3}(p)&=&\left(   \begin{array}{cccc}
p_{L}^{2} &0& -p_{1}p_{3}& -p_{1}p_{0} \\
0&p_{L}^{2}& -p_{2}p_{3}& -p_{2}p_{0} \\
-p_{1}p_{3}& -p_{2}p_{3}& p_{\perp}^{2} &0 \\
-p_{1}p_{0}& -p_{2}p_{0}&0& p_{\perp}^{2}
\end{array} \right).
\nonumber
\eeq

We next determine $a_{1}$, $a_{2}$ and $a_{3}$ such that the difference
\beq
S_{\rm diff}=\int_{\tilde {\mathbb P}}\frac{d^{4}q}{(2\pi)^{4}}\,{\rm Tr}\,{\tilde A}(-p)^{T}M_{\rm diff}(p)
{\tilde A}(p)=
\int_{\tilde {\mathbb P}}\frac{d^{4}q}{(2\pi)^{4}}\,{\rm Tr}\,{\tilde A}(-p)^{T}
\sum_{j=3}^{6}\Pi^{j}(p){\tilde A}(p)\;-\;S_{\rm quad} \label{diff}
\eeq
is maximally non-gauge invariant. By this we mean that the projection of
tensor $M_{\rm diff}(p)$ to a gauge-invariant expression: 
\beq
\left( 1\!\!{\rm l}-\frac{p\,p^{T}}{p^{T}p} \right)M_{\rm diff}(p)
\left( 1\!\!{\rm l}-\frac{p\,p^{T}}{p^{T}p} \right)\;, \nonumber
\eeq
has no local part. This gives a precise determination of $S_{\rm diff}$, which is proportional
to the counterterm to be subtracted. To carry this procedure out, we break
up the second term of (\ref{p-t}) into  a linear combination of $M_{1}$, $M_{2}$ and
$M_{3}$ and a diagonal matrix:
\beq
\sum_{j=3}^{6}\Pi^{j}(p)\!\!&\!\!=\!\!&\!\! 
\frac{1}{192\pi^{2}}\ln \frac{\Lambda}{\tilde \Lambda}\; (1\!\!{\rm l}
-pp^{T})
\nonumber \\ 
&+&\frac{\ln {\tilde b}}{64 \pi^{2}}\left[
\frac{7}{12}M_{1}(p)-\frac{13}{12}M_{2}(p)+\frac{7}{4}M_{3}(p)
\right] \nonumber \\ 
&+&
\frac{\ln {\tilde b}}{64 \pi^{2}}\left(   \begin{array}{cccc}
-\frac{3}{4}p_{\perp}^{2}-\frac{17}{6}p_{L}^{2}    &0&0&0 \\
0&  -\frac{3}{4}p_{\perp}^{2}-\frac{17}{6}p_{L}^{2}   &0&0 \\
0&0& -\frac{17}{12}p_{\perp}^{2}+\frac{7}{4}p_{L}^{2} &0 \\ 
0&0&0&-\frac{17}{12}p_{\perp}^{2}+\frac{7}{4}p_{L}^{2} 
\end{array}
\right)\;.\label{p-t1}
\eeq
The diagonal matrix is maximally non-gauge-invariant. It is local, ${\rm O}(2)\times {\rm O}(2)$
invariant and of dimension four; we
remove it with local counterterms, rendering our ellipsoidal cut-offs gauge invariant, to one 
loop. Therefore 
\beq
a_{1}=\frac{\ln {\tilde b}}{64\pi^{2}}\cdot \frac{7}{12}\;,\;\;
a_{2}=-\frac{\ln {\tilde b}}{64\pi^{2}}\cdot \frac{13}{12}\;,\;\;
a_{3}=\frac{\ln {\tilde b}}{64\pi^{2}}\cdot \frac{7}{4}\;. \nonumber
\eeq

Removing the last term from (\ref{p-t1}) leaves us
with our final result for the polarization tensor
\beq
{\hat \Pi}(p)\!\!&\!\!=\!\!&\!\! \sum_{j=3}^{6}\Pi^{j}(p)-\frac{\ln {\tilde b}}{64 \pi^{2}}\left(   \begin{array}{cccc}
-\frac{3}{4}p_{\perp}^{2}-\frac{17}{6}p_{L}^{2}    &0&0&0 \\
0&  -\frac{3}{4}p_{\perp}^{2}-\frac{17}{6}p_{L}^{2}   &0&0 \\
0&0& -\frac{17}{12}p_{\perp}^{2}+\frac{7}{4}p_{L}^{2} &0 \\ 
0&0&0&-\frac{17}{12}p_{\perp}^{2}+\frac{7}{4}p_{L}^{2} 
\end{array}
\right)\nonumber \\ 
\!\!&\!\!=\!\!&\!\! \frac{1}{192\pi^{2}}\ln \frac{\Lambda}{\tilde \Lambda}\; (1\!\!{\rm l}
-pp^{T})+\frac{\ln {\tilde b}}{64 \pi^{2}}\left[
\frac{7}{12}M_{1}(p)-\frac{13}{12}M_{2}(p)+\frac{7}{4}M_{3}(p)
\right] \;. 
\nonumber
\eeq
One of the terms to be induced in the renormalized action by integrating out
fast degrees of freedom is thereby
\beq
\frac{1}{2}\langle S_{\rm I}\rangle\!\!&\!\!-\!\!&\!\!\frac{1}{4}(\langle S_{\rm I}^{2}\rangle-\langle S_{\rm I}\rangle^{2}) =C_{N}\int_{\tilde {\mathbb P}} \frac{d^{4}p}{(2\pi)^{4}}\;
{\tilde A}^{b}_{\mu}(-p){\tilde A}^{b}_{\nu}(p)\; {\hat \Pi}_{\mu \nu}(p) \nonumber \\
\!\!&\!\!=\!\!&\!\!
C_{N}\int_{\tilde {\mathbb P}} \frac{d^{4}p}{(2\pi)^{4}}\;
{\tilde A}^{b}_{\mu}(-p){\tilde A}^{b}_{\nu}(p)\left\{
\frac{1}{192\pi^{2}}\ln \frac{\Lambda}{\tilde \Lambda}\; (1\!\!{\rm l}
-pp^{T})+\frac{\ln {\tilde b}}{64 \pi^{2}}\left[
\frac{7}{12}M_{1}(p)-\frac{13}{12}M_{2}(p)+\frac{7}{4}M_{3}(p)
\right] 
\right\}\;. \label{polarization-contrib}
\eeq
The other term induced by this integration, namely
$-(\langle S_{\rm II}^{2}\rangle-\langle S_{\rm II}\rangle^{2})/2$, will be discussed next.

We showed in Section 2 that the term $-(\langle S_{\rm II}^{2}\rangle-\langle S_{\rm II}\rangle^{2})/2$ is given by
(\ref{F-F}). This term 
may be expanded in powers of $p$ as
we did for the spherical case in (\ref{expanded-F-F}). The result is
\beq
-\frac{1}{2}(\langle S_{\rm II}^{2}\rangle-\langle S_{\rm II}\rangle^{2}) 
&=&-\frac{C_{N}}{2}\int_{\tilde {\mathbb P}} \frac{d^{4}p}{(2\pi)^{4}} 
{\tilde F}^{b}_{\mu \nu}(-p)
{\tilde F}^{b}_{\mu \nu}(p)
\int_{{\mathbb S}} \frac{d^{4}q}{(2\pi)^{4}} \frac{1}{(q^{2})^{2}}
+\cdots  \nonumber \\
&=&-C_{N}\left[
\frac{1}{16\pi^{2}}\;\ln \frac{\Lambda}{{\tilde \Lambda}}-\frac{b \ln b}{32\pi^{2}(b-1)}
+\frac{{\tilde b} \ln {\tilde b}}{32\pi^{2}({\tilde b}-1)}
\right]
\;\int_{\tilde {\mathbb P}} \frac{d^{4}p}{(2\pi)^{4}} 
{\tilde F}^{b}_{\mu \nu}(-p)
{\tilde F}^{b}_{\mu \nu}(p)
+\cdots 
\;. \label{expanded-F-F1}
\eeq
For $b=1$ and to leading order in $\ln {\tilde b}$, (\ref{expanded-F-F1}) becomes
\beq
-\frac{1}{2}(\langle S_{\rm II}^{2}\rangle-\langle S_{\rm II}\rangle^{2}) 
=-C_{N}\left(
\frac{1}{16\pi^{2}}\;\ln \frac{\Lambda}{{\tilde \Lambda}}
+\frac{\ln {\tilde b}}{32\pi^{2}}\right)
\;\int_{\tilde {\mathbb P}} \frac{d^{4}p}{(2\pi)^{4}} 
{\tilde F}^{b}_{\mu \nu}(-p)
{\tilde F}^{b}_{\mu \nu}(p)
+\cdots 
\;. \label{expanded-F-F2}
\eeq

Putting together (\ref{polarization-contrib}) and (\ref{expanded-F-F2}) gives the following
expression for the new action ${\tilde S}=\int d^{4}x {\tilde {\mathcal L}}$:
\beq
{\tilde {\mathcal L}}\!\!&\!\!=\!\!&\!\!
\frac{1}{4}\left(\frac{1}{g_{0}^{2}}-\frac{11C_{N}}{48\pi^{2}}\ln\frac{\Lambda}{\tilde \Lambda}
-\frac{C_{N}\ln{\tilde b}}{64\pi^{2}} \right)\left(
{\tilde F}_{01}^{2}+{\tilde F}_{02}^{2}+{\tilde F}_{13}^{2}+{\tilde F}_{23}^{2}
\right) 
+\frac{1}{4}\left(\frac{1}{g_{0}^{2}}-\frac{11C_{N}}{48\pi^{2}}\ln\frac{\Lambda}{\tilde \Lambda}
-\frac{37C_{N} \ln{\tilde b}}{192\pi^{2}} \right){\tilde F}_{03}^{2}
\nonumber \\
\!\!&\!\!+\!\!&\!\!
\frac{1}{4}\left(\frac{1}{g_{0}^{2}}-\frac{11C_{N}}{48\pi^{2}}\ln\frac{\Lambda}{\tilde \Lambda}
-\frac{17C_{N} \ln{\tilde b}}{192\pi^{2}} \right){\tilde F}_{12}^{2}+\cdots\;.
\label{renormalized-action}
\eeq

\section{The longitudinally rescaled Yang-Mills action}

\setcounter{equation}{0}
\renewcommand{\theequation}{5.\arabic{equation}}

The main result of Section 4, equation (\ref{renormalized-action}), tells what happens after
aspherically integrating out degrees of freedom. We will write this in a form which
allows comparison with standard renormalization with an isotropic cut-off, {\em i.e.}
(\ref{asymptotic-freedom}). We define ${\tilde g}_{0}$ using (\ref{asymptotic-freedom}). To
leading order in $\ln {\tilde b}$, the effective coupling in the first term of (\ref{renormalized-action}) is
given by
\beq
\frac{1}{g_{\rm eff}^{2}}=\frac{1}{g_{0}^{2}} 
-\frac{11C_{N}}{48\pi^{2}} \ln \frac{\Lambda}{\tilde \Lambda}
-\frac{C_{N}\ln {\tilde b}}{64\pi^{2}}
=\frac{1}{{\tilde g}_{0}^{2}} \,{\tilde b}^{-\frac{C_{N}}{64\pi^{2}}{\tilde g}_{0}^{2}}
+\cdots\;. \nonumber
\eeq
After we set ${\tilde b}=\lambda^{-2}$,
we find to leading order in $\ln \lambda$ 
\beq
g_{\rm eff}^{2}={\tilde g}_{0}^{2}\,\lambda^{-\frac{C_{N}}{32\pi^{2}}{\tilde g}_{0}^{2}} \;.
\label{eff-coupling}
\eeq
and
\beq
{\tilde {\mathcal L}}=\frac{1}{ 4g_{\rm eff}^{2} }
\,{\rm Tr}\,\left(
{\tilde F}_{01}^{2}+{\tilde F}_{02}^{2}+{\tilde F}_{13}^{2}+{\tilde F}_{23}^{2}
+\lambda^{\frac{17C_{N}}{48\pi^{2}}{\tilde g}_{0}^{2}}{\tilde F}_{03}^{2}+
\lambda^{\frac{7C_{N}}{48\pi^{2}}{\tilde g}_{0}^{2}}{\tilde F}_{12}^{2}
\right)+\cdots \;,
\nonumber
\eeq
where the corrections are of order $(\ln \lambda)^{2}$. We perform the
rescaling of longitudinal coordinates, $x^{L}\rightarrow \lambda x^{L}$, drop the tildes
on the fields, and Wick-rotate back to Minkowski signature, to find the 
longitudinally-rescaled effective Lagrangian
\beq
{\mathcal L}_{\rm eff}=\frac{1}{ 4g_{\rm eff}^{2} }
\,{\rm Tr}\,\left(
{F}_{01}^{2}+{F}_{02}^{2}-{F}_{13}^{2}-{F}_{23}^{2}
+\lambda^{-2+\frac{17C_{N}}{48\pi^{2}}{\tilde g}_{0}^{2}}{F}_{03}^{2}-
\lambda^{2+\frac{7C_{N}}{48\pi^{2}}{\tilde g}_{0}^{2}}{F}_{12}^{2}
\right)+\cdots
\;,\label{effective-lag}
\eeq
where again the corrections are of order $(\ln \lambda)^{2}$. Comparing this with the classically-rescaled action (\ref{action}) we see that
the field-strength-squared terms are anomalously rescaled.

If we naively consider  the limit as  
$\lambda\rightarrow 0$ of (\ref{effective-lag}), all couplings become zero or infinite, except 
$g_{\rm eff}$ \cite{Verlinde-squared}.  For very high energy, that is for small $\lambda$, this effective coupling becomes strong, as can immediately
be seen from (\ref{eff-coupling}). We are fortunate, however, that this energy is
enormous. If we take ${\tilde g}_{0}$ of order one, then 
\beq
g_{\rm eff}^{2}\sim \lambda^{-\frac{1}{100}} \;.
\eeq 
This tells us that $g_{\rm eff}^{2}$ is less than a number of order ten, unless 
$\lambda$ is roughly less than an inverse googol,
$\lambda\sim 10^{-100}$. Thus the experimentally
accessible value of  $g_{\rm eff}$ is small. Even so, there 
is the concern that
the coefficient of $F_{12}^{2}$ in the effective Lagrangian is very 
small as $\lambda\rightarrow 0$. This 
is also true
for the classically-rescaled theory (\ref{action}) \cite{mypaper}. This means that
there is very little energy in longitudinal magnetic flux. Hence the
longitudinal magnetic flux fluctuates wildly. If we call the coefficient
of this
term in the Lagrangian $1/(4g_{L}^{2})$, then
\beq
g_{L}^{2}=g_{\rm eff}^{2}\lambda^{-2-\frac{7C_{N}}{48\pi^{2}}{\tilde g}_{0}^{2}}\;, \label{long-coupl}
\eeq
which explodes for small $\lambda$, even if $g_{\rm eff}$ is small.

\section{Implications for effective high-energy theories}

\setcounter{equation}{0}
\renewcommand{\theequation}{6.\arabic{equation}}

We have determined how a quantized non-Abelian gauge action changes
under a longitudinal rescaling $\lambda <1$, but $\lambda \approx 1$. Though
our analysis suggests the form of the effective 
action for the high-energy limit, $\lambda \ll1$, we cannot prove that this
form is correct. The 
main problem concerns
how the Yang-Mills action changes
as $\lambda$ is decreased. The coefficient of the 
longitudinal magnetic field squared, in the action, decreases, as $\lambda$ is
decreased. Eventually, we can no longer compute how couplings will run.

Our difficulty is very similar to that of finding the spectrum of a non-Abelian gauge
theory. Assuming that there is no infrared-stable fixed point
at non-zero bare coupling, a guess for the long-distance effective theory
is a strongly-coupled cut-off action. The regulator can be a lattice, for example. One
can then use strong-coupling expansions to find the spectrum. The
problem is that no one knows how to specify the true cut-off theory (which 
presumably has many terms, produced by integrating over all the short-distance degrees of
freedom). The best we can do is guess the regularized strongly-coupled
action. Such
strong-coupling theories
are not (yet) derivable from 
QCD, but are best thought of as models of the strong 
interaction at large distances.

Similarly, we believe that
(\ref{action}) for $\lambda \ll1$, and variants 
we discuss below, cannot be proved to describe 
the strong interaction at high energies. Thus it appears that the same statement applies
to the the BFKL/BK theory (designed to describe the region where Mandelstam variables 
satisfy $s\gg t\gg \Lambda_{QCD}$)
\cite{BalitskiFadKurLip}, \cite{BK}. Two
closely-related problems in this theory are lack of unitarity and infrared diffusion
of gluon virtualities. These problems indicate that the BFKL theory breaks down
at large length scales. There is numerical evidence \cite{stasto} that unitarizing using the
BK evolution equation \cite{BK}  suppresses diffusion into the infrared and
leads to saturation,  at least for fixed small impact parameters. This
BK equation is a non-linear generalization of the BFKL evolution equation. The non-linearity
only becomes important at small $x$, at large longitudinal distances, where perturbation
theory is not trustworthy.

In the color-glass-condensate picture \cite{McLerranVenugopalan}, \cite{CGC}, the Yang-Mills
action with $\ln \lambda =0$ is coupled to
sources. The classical field strength is purely transverse. If this action is quantized, however,
this is no longer the case. The
fluctuations of the longitudinal magnetic field ${\mathcal B}_{3}$ will become extremely
large (this can be seen by inspecting (\ref{action}) and (\ref{ContHamiltonian})). In principle, we
would hope to derive the color-glass condensate by a longitudinal renormalization-group
transformation, with background sources. The obstacle to doing this is precisely the problem
of large fluctuations of ${\mathcal B}_{3}$. This does not suggest any inconsistency of the 
color-glass-condensate idea itself, but indicates how difficult it may be to
establish the color-glass condensate directly in QCD.

Finally we wish to comment on an approach to soft-scattering and total cross sections. In Reference
\cite{mypaper} an effective lattice 
SU(N) gauge theory was proposed. This gauge theory is a regularization of
(\ref{ContHamiltonian}) and (\ref{Gauss}). This gauge theory can be formulated
as coupled $(1+1)$-dimensional
${\rm SU}(N)\times {\rm SU}(N)$ nonlinear sigma models and
reduces to a lattice Yang-Mills theory at $\lambda=1$ (in which
case, it is equivalent to the light-cone lattice theory of
Bardeen et. al. \cite{Bardeen}). The nonlinear sigma model
is asymptotically free and has a mass gap. These facts together with the assumption
that the terms proportional to $\lambda^{2}$ are a weak perturbation leads to
confinement and diffraction in the gauge theory. Similar gauge
models in $(2+1)$ dimensions were proposed
as laboratories of color confinement \cite{PhysRevD71}, and string tensions for
different representations \cite{PhysRevD75-1}, the low-lying glueball 
spectrum \cite{PhysRevD75-2}, and corrections of higher order in order $\lambda$ to the
string tension  \cite{PhysRevD74} were found (these calculations were performed using
the exact S-matrix \cite{abda-wieg} and form factors \cite{KarowskiWeisz} of
the $(1+1)$-dimensional nonlinear sigma model). In such
theories (whether in $(2+1)$ or $(3+1)$ 
dimensions), transverse electric flux is built of massive partons (made entirely
of glue, but not conventional gluons). These
partons move (and scatter)
only longitudinally, to leading order in $\lambda$. The behavior
of the such gauge-theory models
is very close to the picture of the forward-scattering 
amplitude suggested by Kovner \cite{Kovner}.

The effective gauge theory of Reference \cite{mypaper} has a small
value of  $g_{\rm eff}$, as well as a small value of $\lambda$ ,in the Hamiltonian 
(\ref{ContHamiltonian}). We have found in Section 5 that $g_{\rm eff}$ grows extremely
slowly, as the energy is increased. If we can naively extrapolate our results to extremely
high energies, this effective gauge theory appears correct. We should not, however, regard this as  
proof that the effective theory is valid, since the perturbative calculation of Section 5 breaks
down at such energies.

\section{Discussion}
\setcounter{equation}{0}
\renewcommand{\theequation}{7.\arabic{equation}}

In this paper, we found how the action of an SU($N$) gauge changes under longitudinal
rescaling at one loop. This was done by a Wilsonian renormalization procedure. As the
energy increases, the
coefficient of $F_{12}^{2}$ in the action eventually becomes too small to trust the method
further. Therefore, neither classical nor perturbative
methods may be entirely trusted beyond a certain energy. The breakdown of these methods 
at high energies is
similar to the breakdown of perturbation theory to compute the force between
charges at large distances, in an 
asymptotically-free theory. Nonetheless, high-energy 
effective theories, inspired by the longitudinal-rescaling idea,
may be phenomenologically useful.

The next step is to repeat our calculation including Fermions. Aside from the importance  
of considering QCD with quarks, it would be interesting to study how longitudinal
rescaling affects the QED action.

We should point out that another way to derive our effective Lagrangian (\ref{effective-lag})
and investigate anomalous dimensions of other operators
would be to carefully study Green's functions of the operator
\beq
{\mathcal D}(x)=x^{0}{\mathcal T}_{00}(x)+x^{3}{\mathcal T}_{03}(x) \;,\label{rescaling-op}
\eeq
where ${\mathcal T}_{\mu \nu}(x)$ is the stress-energy-momentum tensor. The 
spacial integral of this operator generates longitudinal rescalings on states. Correlators
of products of ${\mathcal D}(x)$ and other operators 
could be studied with simpler regularization methods (such as
dimensional regularization) instead of our
sharp momentum cut-off. The commutator of ${\mathcal D}(x)$ and an operator
${\mathcal O}(y)$ will reveal how ${\mathcal O}(y)$ behaves under longitudinal rescaling. Such an
analysis should be easier than the method used in this
paper, especially beyond one loop.

\begin{acknowledgments}

P.O. had early conversations about this work with 
Poul Henrik Damgaard at the Niels Bohr International Academy and Gordon Semenoff 
at the University of British Columbia. We would also like to thank Adrian Dumitru and
Jamal Jalilian-Marian for  extensive discussions. Finally, we are
grateful to Rob Pisarski for making a  
helpful technical suggestion. 
This research was supported in
part by the National Science Foundation, under Grant No. PHY0653435
and by a grant from the PSC-CUNY.

\end{acknowledgments}

\newpage

\begin{figure}[ht]

\begin{picture}(80,100)(150,100)

\linethickness{0.5mm}

\put(0,60){\put(-15,0){\vector(1,0){100}}

\put(100,-5){$x^{L}$}

\put(-15,0){\vector(0,1){100}}

\put(-20, 105){$x^{\perp}$}}

\linethickness{1mm}

\put(20, 80){\line(0,1){100}}
\put(50, 80){\line(0,1){100}}
\put(80, 80){\line(0,1){100}}

\put(0,160){\line(1,0){100}}
\put(0,130){\line(1,0){100}}
\put(0,100){\line(1,0){100}}

\put(120,125){$\implies$}
\put(117,110){Block!}

\put(180, 80){\line(0,1){100}}

\put(210, 80){\line(0,1){5}}
\put(210, 90){\line(0,1){5}}
\put(210, 100){\line(0,1){5}}
\put(210, 110){\line(0,1){5}}
\put(210, 120){\line(0,1){5}}
\put(210, 130){\line(0,1){5}}
\put(210, 140){\line(0,1){5}}
\put(210, 150){\line(0,1){5}}
\put(210, 160){\line(0,1){5}}
\put(210, 170){\line(0,1){5}}
\put(210, 180){\line(0,1){5}}

\put(210, 80){\line(0,1){5}}

\put(240, 80){\line(0,1){100}}

\put(160,160){\line(1,0){100}}
\put(160,130){\line(1,0){100}}
\put(160,100){\line(1,0){100}}

\put(280,125){$\implies$}
\put(278,110){Scale!}

\put(340, 80){\line(0,1){100}}
\put(370, 80){\line(0,1){100}}
\put(400, 80){\line(0,1){100}}

\put(320,160){\line(1,0){100}}
\put(320,130){\line(1,0){100}}
\put(320,100){\line(1,0){100}}

\end{picture}

\end{figure}

\vspace{100pt}

\noindent
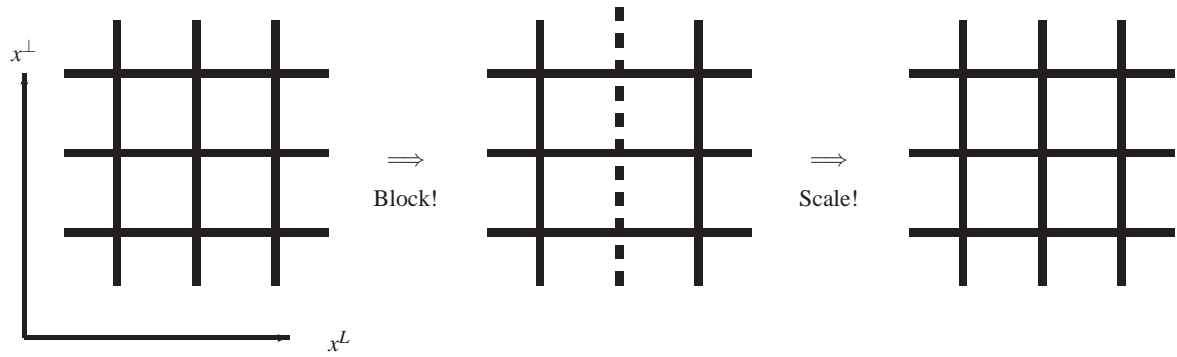
FIG.1. Rescaling of field theory 
on a lattice with $\lambda=1/2$. First, a Kadanoff transformation increases the 
longitudinal
lattice spacing. The spacing is
then restored to its original value by a longitudinal rescaling

\end{document}